\documentstyle[amssymb,twocolumn,prl,aps]{revtex}

\begin{document}

\title{Field-Induced Crossover and Colossal Magnetoresistance in La$_{0.7}$Pb$_{0.3} $MnO$_{3}$}
\author{Y. Y. Xue$^{1}$, B. Lorenz$^{1}$, A. K. Heilman$^{1}$, M. Gospodinov$^{2}$,
S. G. Dobreva$^{2}$ and C. W. Chu$^{1,3}$}
\address{$^{1}$Texas Center for Superconductivity, University of Houston, Houston TX 77204\\
$^{2}$Laboratory of Crystal Growth, Institute of Solid State Physics,\\
Bulgarian Academy of Sciences 72 Tzarigradsko Chaussee Blvd., Sofia Bulgaria 1784\\
$^{3}$Lawrence Berkeley National Laboratory, 1 Cyclotron Road, Berkeley California 94720}
\date{\today}
\maketitle

\begin{abstract}
A field-induced crossover is observed in the resistivity ($\rho $) and
magnetization (M) of a La$_{0.7}$Pb$_{0.3}$MnO$_{3}$ single crystal. The
field-dependence of $\rho $ and M suggests that a small spin-canted species
with mean-field-like interactions dominates at low fields (H), whereas, individual
spins and 3D Ising/Heisenberg models describe the high-H behavior rather
well. Around the ferromagnetic transition, an H-induced destruction of the
small spin-canted magnetic polarons is accompanied by large magnetoresistance.
\end{abstract}

Colossal magnetoresistance (CMR), a large resistivity ($\rho $)
drop induced by the magnetic field (H) in A$_{1-x}$A'$_{x}$MnO$_{3}$ (where
A = rare earth elements and A'= divalent cations such as Ba, Sr, Ca or Pb),
has attracted great attention recently \cite{ramirez}. It is widely
recognized that mean-field models with the double exchange (DE) and the
Jahn-Teller (JT) distortion can not quantitatively describe CMR \cite
{millis,roder}. Various phase-segregation models, which treat manganites as
metallic ferromagnet-clusters embedded in an insulating paramagnet-matrix,
were subsequently proposed \cite{good,teresa,moreo}.\ According to these
models, CMR results from the formation of percolating paths between these
clusters, whose concentration and size increase with H. This proposition is
consistent with various observations in manganites with small (A, A') \cite
{teresa,lanzara,uehara,fath}, where the large size-mismatch between (A, A')
and Mn favors mesoscopic charge-segregation. Such a percolating description,
however, may not be proper for manganites with good size-match and optimal
doping (whose resistivity is metallic above T$_{C}$), although experimental
data in these compounds, {\it e.g.} the pair-density-function of La$_{1-x}$Sr%
$_{x}$MnO$_{3}$ \cite{louca} and the non-Curie-Weiss susceptibility ($\chi $%
) in La$_{1-x}$Pb$_{x}$MnO$_{3}$ \cite{leung}, strongly suggest the
existence of local magnetic structures. The role of these structures in the
CMR is still unknown. This uncertainty is the motivation behind the present
investigation.

In this work, the bulk-magnetization (M) and resistivity ($\rho $) of a La$%
_{0.7}$Pb$_{0.3}$MnO$_{3}$ single-crystal are measured. An H-induced
crossover is observed at a threshold field H$_{T}$(T) with the following
properties: 1) $\rho $ is independent of both H and M for H\ 
\mbox{$<$}%
\ H$_{T}$, but decreases exponentially with M for H far above H$_{T}$; 2) $%
\partial \rho $(M,T)/$\partial $T peaks at T$_{C}$ as that predicted in
critical scaling only if H\ 
\mbox{$<$}%
\ H$_{T}$, but becomes zero with a universal $\rho $ $\propto $ exp(-M/M$_{%
\text{o}}$) at higher fields; 3) In the critical region, M(T, H) follows a
mean-field-like scaling below H$_{T}$, but switches to that of
3D-Heisenberg/Ising models far above; 4) The low-field Curie constant, after
corrections for the critical fluctuations, is three times larger than that
of isolated Mn$^{3+}$/Mn$^{4+}$'s. The data, therefore, suggest a robust but
small magnetic-structure with canted spins, similar to the one-site small
polarons observed in La$_{1-x}$Sr$_{x}$MnO$_{3}$ \cite{louca}. We propose
that the CMR in these less-distorted manganites is closely related to an
H-induced destruction of the short-range spin-canted correlations.

Large single crystals of La$_{0.7}$Pb$_{0.3}$MnO$_{3}$ have been grown in
sealed Pt crucibles by slowly cooling molten mixtures of La$_{2}$O$_{3}$
(99.99\%, Aldrich), MnO$_{2}$ (Spec. pure), PbO (Spec. pure), and PbF$_{2}$
(99.99\%, Johnson- Mathey). A sample of 2$\times $2$\times $5 mm$^{3}$ was
cut from a crystal and used here. The measurements of M and $\rho $ were
made inside a Quantum Design 5-T SQUID magnetometer, and were carried out
with the field increasing from 0.01 to 5 T stepwise at a fixed temperature.
Care has been taken to reduce temperature fluctuations and field relaxation,
and the H was calculated based on the demagnetization factor obtained from
the low-field dM/dH at T$_{C}$.

X-ray diffraction analysis at room temperature shows a cubic symmetry of
Pm3m without noticeable orthorhombic or rhombohedral distortions. The
lattice parameters are {\it a} = {\it b} = {\it c} = 3.8938(4) \AA\ and $%
\alpha $ = $\beta $ = $\gamma $ = 90$^{\text{o}}$. Both the symmetry and the
lattice parameter demonstrate an ion-size match better than that in La$_{0.7}
$Sr$_{0.3}$MnO$_{3}$, although the related lattice-disorder (due to the
larger size-difference between La and Pb) seems to suppress its T$_{C}$
slightly \cite{hwang}. The composition was analyzed using a JEOL JXA 8600
electron microprobe. A stoichiometry La:Pb:Mn:O =
0.679(9):0.303(6):1:3.06(5) is observed over the whole surface of the
crystal. The composition spread is well within our instrumental resolution
of 1-3\% for cations and $\approx $\ 5\% for oxygen.

The $\rho $(T,M) observed is shown in figure 1, with the corresponding raw
data shown in the inset. Two very different regions can be clearly
identified. At the high field limit, $\rho $ decreases with M almost
universally as $\rho $ $\propto $ exp(-M/M$_{\text{o}}$) (the solid line in
Fig. 1) with a fitting parameter M$_{\text{o}}$ = 23 emu/g $\approx $\ 0.2
T. Below a threshold field H$_{T}$, however, $\rho $ becomes M-independent.
At 320 K, for example, $\rho $ varies less than 0.2\% when M increases to
half of the saturated moment M$_{\text{sat}}$ at 0.14 T. It is interesting
to note that H$_{T}$ should scale with $\varepsilon ^{\beta +\gamma }$(where 
$\varepsilon $, $\beta $ and $\gamma $ are the reduced temperature $%
\varepsilon =$(T-T$_{C}$)/T$_{C}$ and two critical exponents \cite{ghosh})
in scaling models, and a 30-times increase is expected at $\varepsilon $ =
-0.04, -0.004, and 0.04 (T = 320, 332, and 346 K). However, the measured
values, defined as the field where ($\rho $(0)-$\rho $(H$_{T}$))/$\rho $(0)
= 0.01, vary more slowly, {\it i.e.} 0.15, 0.04, and 0.1 T, respectively
(filled triangles in Fig. 1). The crossover, therefore, is unlikely due to
critical fluctuations alone. In fact, various scaling calculations show that
the magnetoresistance below T$_{C}$ should be proportional to either M$^{2}$ 
\cite{millis,fishera} or $\left| \text{H}\right| $ \cite{balberg} up to a
characteristic field of {\it k}$_{B}$T$_{C}$/$\mu $ $\approx $\ 200 T (where 
$\mu $ is the moment of the spins), which is in disagreement with our data.


Previously, similar exponential M-dependence of $\rho $ has been observed in
manganite films \cite{hundley,sun}. This dependence was interpreted by
assuming {\it ln}$\rho $ $\propto $ -{\it t, t} $\propto $ cos($\theta _{ij}$%
/2) and M $\propto $ cos($\theta _{ij}$/2) in the framework of DE, where 
{\it t}{\rm \ }is the transfer-integral and $\theta _{ij}$ is the angle
between adjacent spins. In doing so, an implicit $\theta _{ij}$ = 2$%
\overline{\theta }$ was assumed ($\overline{\theta }$ is the average angle
with H), and all Mn$^{3+/4+}$ spins were treated as independent \cite
{hundley,zener}.

The data, therefore, demonstrate\ that the spins around a carrier in La$%
_{0.7}$Pb$_{0.3}$MnO$_{3}$ are strongly correlated below H$_{T}$. A field of
0.15 T at 320 K, for example, substantially aligns the spins to raise M to $%
\approx $ 65\% of its saturation value M$_{\text{sat}}$, but leaves $\rho $
almost unchanged. The $\overline{\theta }$ = arccos(M/M$_{\text{sat}}$)
decreased with H by more than 40$^{\text{o}}$ while the variation of $\theta
_{ij}$ is negligible. This demonstrates a short-range spin-correlation. The
spins within a mean free path ($\approx $ 10 \AA\ \cite{schiffer}) from a
carrier have to rotate as a whole under the field, {\it i.e.} a robust
short-range spin-correlation.

It should be pointed out that the M-increase is not caused by domain
rotations. The second-order transition of La$_{0.7}$Pb$_{0.3}$MnO$_{3}$,
which is expected from its lattice match \cite{moutis} and demonstrated in
our scaling analysis, eliminate domains, which mark the discontinuity of
order parameters. The residual ones, if any, should be isolated around
defects, and involve only a small fraction of spins. Experimentally, the
H-induced nucleation of domains around T$_{C}$\ should appear as a jump-like
(or S-shape) feature at low H in the Arrott plot \cite{moutis}, but could
not be observed in our data (Fig.2b).

This H-independent $\rho $ at low-fields drops quickly with the decrease of
temperature, and d$\rho $/dT $\equiv $ $\partial \rho $/$\partial $T$\mid
_{M}$ peaks at T$_{C}$ = 334(1) K. A scaling fit of d$\rho $/dT $\propto
\left| \varepsilon \right| ^{-\alpha }$ then leads to $\alpha $ = 0.03(5)
within $\left| \varepsilon \right| $ $\leq $ 0.005. Although the
uncertainties in both T$_{C}$ and $\alpha $ are large, the low-field $\rho $
seems to follow the same scaling law as that in Ni. At high H, however, $%
\partial \rho $/$\partial $T$\mid _{M}=0$, and the $\rho $ depends on M
universally as reported previously \cite{hundley,sun}. An H-induced
crossover, therefore, is clear in the T-dependence of $\rho $ also.

The data demand a robust canted-FM spin-correlation around the carriers
below H$_{T}$. In fact, the resistivity drops by 50\% or more when the field
increases from H$_{T}$ to 5 T, which requires a large $\theta _{ij}$
(estimated $\approx $ 10-40$^{\text{o}}$ in the temperature range
investigated) below H$_{T}$. The fact that these magnetic structures are
destroyed by a H $\gg $ H$_{T}$, as shown in $\rho $(T,M) discussed above
and the M(T,H) below, supports the conclusion. It is interesting to note
that the pair-density function (PDF) data in similar La$_{1-x}$Sr$_{x}$MnO$%
_{3}$ show that the holes form one-site small polarons at 300-350 K, but as
moving carriers at 10 K, although both temperatures are below the corresponding T$%
_{C}$ \cite{louca}. This can happen in the DE model only if
the $\theta _{ij}$ between the hole (Mn$^{4+}$) and the adjacent Mn$^{3+}$
is large enough to make the hopping time longer than the time-window of PDF.
A spin-canting with a T-dependent $\theta _{ij}$ will offer a natural
interpretation for the unusual PDF data. The $\rho $ and M data here,
therefore, present an indirect evidence for spin canting, which was not
observed before \cite{moreo}. Detailed neutron diffraction investigations
are needed to explore this issue.

A canted-FM correlation may occur if there is competition between various
magnetic interactions, as expected in early models \cite{zener}. This view,
however, has been challenged recently \cite{moreo,golosov}. It has been
pointed out \cite{moreo} that mechanisms other than those of de Gennes \cite
{zener} may be needed under certain conditions if canted states exist. Spin
canting, in a sense, is incompatible with mesoscopic segregation between FM
and AFM regions, and would not be expected in highly distorted manganites.
Some two-orbital model calculations, however, show that phase-segregation
appears in optimally doped manganites only if the JT strength ($\lambda $%
)\thinspace\ is large enough \cite{moreo}. The almost perfect structure of La%
$_{0.7}$Pb$_{0.3}$MnO$_{3}$, therefore, may eliminate mesoscopic charge
segregations and stabilize the spin canting.

In order to further explore the magnetic structures, the bulk magnetization
is analyzed in the critical region using the scaling function: 
\begin{equation}
(H/M)^{1/\gamma }=a\cdot \varepsilon +b\cdot M^{1/\beta }  \eqnum{1}
\end{equation}
where {\it a} and {\it b} are two critical amplitudes, and $\beta $ and $%
\gamma $ are the critical exponents defined before. It is known that the
values of $\beta $ and $\gamma $ depend on the range of interactions \cite
{fisher}. They will follow the mean-field theory with $\beta $ $\approx $
0.5 and $\gamma $ $\approx $ 1 for a long range interaction, but
3D-Heisenberg or -Ising models with{\it \ }$\beta $ $\approx $ 0.33 and $%
\gamma $ $\approx $ 1.3 for a nearest-neighbor one \cite{fisher}.
Previously, the critical phenomena have been investigated \cite
{ghosh,moutis,mohan} with rather divergent results. Sample differences were
suggested as the main reason for the discrepancy \cite{ghosh}. We propose
that the crossover discussed above may play a major role. The magnetization
data between -0.04 $\leq $ $\varepsilon $ $\leq $ 0.04 was, therefore,
analyzed separately above and below 1 T in Fig. 2 and its Inset,
respectively. The field of 1 T, which is shown in the figures as filled
diamonds, is chosen between H$_{T}$ and the field where $\rho $ starts to
depend on M exponentially. The parameters $\beta $ and $\gamma $ in Eq. 1
were first adjusted so that the isotherms in the Arrott plots of M$^{1/\beta
}$ vs. (H/M)$^{1/\gamma }$, form a system of parallel straight lines (Fig.
2) \cite{arrott,kaul}. The intercepts Y = M$_{S}$($\varepsilon $) and X = 1/$%
\chi $%
\mbox{$\vert$}%
$_{H=0}$ were deduced, and the Kouvel-Fisher (KF) equations of Y/(dY/dT) =
(T-T$_{C}$)/$\beta $ and X/(dX/dT) = (T-T$_{C}$)/$\gamma $ plotted against T 
\cite{kouvel}. The intercepts of the KF plots were then used as T$_{C}$, and
their slopes as 1/$\beta $ and 1/$\gamma $, respectively. Finally, the
scaling law M= H$^{1/\delta }$ at T$_{C}$, the constraint $\delta $ -1-$%
\gamma $/$\beta $ = 0, as well as the scaling hypothesis of M(H,T)/$%
\varepsilon ^{\beta }$ = {\it f}$_{\pm }$[H$\cdot \varepsilon ^{-\beta
-\gamma }$] were used to verify the scaling over the field ranges specified.
These constraints are satisfied over the field range specified within the
data uncertainty, which is the standard deviation from three consecutive
measurements. Two sets of rather different fitting parameters are indeed
obtained. T$_{C}$ = 334.4(1) K, $\beta $ = 0.33(1), and $\gamma $ = 1.27(2)
(Fit A) were obtained for M(H $\geq $ 1 T) (Fig. 2). The M(H $\leq $ 1 T)
observed, however, significantly differs from this Arrott plot (Fig. 2).
While the low-field deviations below a few hundred Oe, such as those in the
Inset of Fig. 2, has been observed in most ferromagnets due to some
yet-to-be-found reasons \cite{kaul}, those up to 1 T are rather unusual. The
M(H $\leq $ 1 T), on the other side, can be fit well with T$_{C}$ = 336.5(2)
K, $\beta $ = 0.50(2), and $\gamma $ = 1.0(1) (Fit B) (Inset, Fig. 2). It is
interesting to note that the critical exponents of Fit A are very close to
those predicted by 3D Ising/Heisenberg models, and those of Fit B are in
good agreement with the mean-field-model predictions. This seems to be
consistent with the local structures proposed above. With larger physical
sizes, the interaction between the magnetic polarons may be mean-field like,
but the 3D-Heisenberg/Ising models will be more proper when the species are
individual spins. We, therefore, associate the unusual scaling behavior with
the crossover observed in $\rho $(T,M). It should also be noted that the Fit
A is similar to the previous result in La$_{0.7}$Sr$_{0.3}$MnO$_{3}$, where
a field up to 5 T was used \cite{ghosh}, and Fit B to that in La$_{0.8}$Sr$%
_{0.2}$MnO$_{3}$, where the measurements were limited to below 1 T \cite
{mohan}. It is possible that the canted spin-correlation and the related
field-induced crossover are common features in the less distorted
manganites, although their net effects may depend on the structure
parameters, {\it e.g. }the tolerance factor and the lattice-disorder \cite
{data}.


The susceptibility $\chi $ = M/H was measured up to 400 K at various H to
estimate the size of the structure. The (T,H) phase-space, where the
canted-spins exists, was first explored. The dM/dH at 340 K and 370 K was
calculated, and fit with the parameters in Fit B (Inset, Fig. 3). While only
the M below 1 T shows reasonable agreement with the fit at 340 K, all M fits
well at 370 K. The field needed to induce the crossover appears to increase
with T, and the spin-canting structure persists above T$_{C}$ at low-fields.
The generalized Curie-Weiss equation of $\chi $ = {\it C}$\varepsilon
^{-\gamma \prime }$/T was then used to estimate the number of the spins
involved, where $\gamma $' and {\it C} are an effective critical exponent
and Curie constant, respectively \cite{arrotta}. This equation is a
high-order expansion of the 3D Heisenberg model, and fits the magnetization
of Ni between 10$^{-3}$ $\leq $ $\varepsilon $ $\leq $\ 0.8 excellently with
the fit parameters consistent with those obtained from other methods \cite
{arrotta,seeger}. The $\chi \mid _{H=0}$ deduced from the Arrott plots of Fig
2 and its inset was used below 340 K, and the M/H directly measured above 
\cite{arrotta,seeger}. The obtained Curie constant corresponds to an effective
moment of 4.5 and 12 $\mu _{B}$ for the Fits A and B, respectively, where $%
\mu _{B}$ is the Bohr magneton. Although an accurate analysis is difficult
due to the limited temperature range, these values suggest that the spin
canting structure involves only the nearest neighbors, in agreement with
that proposed above.


In conclusion, a field induced crossover is observed in La$_{0.7}$Pb$_{0.3}$%
MnO$_{3}$ for both $\rho $(T, M) and M(T, H) around T$_{C}$. Above 1 T, M(T,
H) varies with $\varepsilon $ similar to the predictions of 3D-spin models,
and $\rho $ $\propto $ exp(-M/M$_{o}$) can be accommodated with the DE
model. At lower fields, however, M(T, H) follows mean-field-like
fluctuations, and $\rho $ is independent of M as if the spin alignment
around the holes is no longer affected by H, indicative of a small but
robust canted spin-structure, likely the one-site magnetic polaron
previously reported. Our data suggest that the destruction of the
correlation, which has been ignored in most models, is an important
mechanism in the CMR of the less distorted manganites.

\acknowledgments

We thank Dr Z. Y. Weng for a careful review of the manuscript, D. K. Ross
and R. Bontchev for the microprobe and XRD measurements. The work is
supported in part by NSF Grant No. DMR 9804325, the T. L. L. Temple
Foundation, the John and Rebecca Moores Endowment and the State of Texas
through TCSUH at University of Houston; and by the Director, Office of
Energy Research, Office of Basic Energy Sciences, Division of Materials
Sciences of the U.S. Department of Energy under Contract No.
DE-AC03-76SF00098 at LBL.

\begin{figure}[h]
\caption{$\protect\rho $ vs M at different temperatures. $\blacktriangle $ $%
\protect\rho $ at H$_{T}$; solid line: $\protect\rho \propto $ exp(-M/23).
Inset: the corresponding raw data.}
\label{fig1}
\end{figure}

\begin{figure}[h]
\caption{a) The measured M vs H at various T without the demagnetization
corrections. \ b) The Arrott plot based on M at H $\geq $ 1 T. $%
\blacklozenge $: data at H = 1 T; lines: fit A of Eq. 1. Inset: Arrott plot
for M with H $\leq $ 1 T. $\ \blacklozenge $: data at H = 1 T; lines: fit B
of Eq. 1.}
\label{fig2}
\end{figure}

\begin{figure}[h]
\caption{$\protect\chi $ vs $\protect\varepsilon ^{-1}$. \ Upper curve (3x): 
$\bullet $: $\protect\chi $ at 0.1 T; $\diamondsuit $: extrapolated $\protect%
\chi |_{H=0}$ from M(H$\geq $1 T); solid line: fit with T$_{C}$ = 332.4 K
and $\protect\gamma $' = 1.22. Lower curve: $\blacktriangle $: $\protect\chi 
$ at 0.1 T; octagon: $\protect\chi |_{H=0}$ from M(H$\leq $1 T); solid line:
fit with T$_{C}=$336.5 K and $\protect\gamma $' = 1.04. Inset: dM/dH vs. H.
o: data at 340 K; $\triangle $: data at 370 K; solid lines: scaling fits
from M(H $\leq $1 T).}
\label{fig3}
\end{figure}


\begin{references}
\bibitem{data}  The data of La$_{0.7}$Sr$_{0.3}$MnO$_{3}$ in \cite{ghosh}
did not show deviations in the Arrott plots down to 0.01 T. Note that,
however, a T-induced crossover with the one-site magnetic polarons and an
extended DE-like configuration at 300 and 10 K, respectively, has been
observed by PDF in the same compound \cite{louca}, an H-induced crossover
could also be expected. The value of H$_{T}$ may vary. The better size-match
in La$_{0.7}$Pb$_{0.3}$MnO$_{3}$, for example, might be the factor for a
larger H$_{T}$ here.
\end{references}
\end{document}